\documentclass[aps,prl,floatfix,twocolumn,showpacs]{revtex4}
\usepackage{epsfig}
\usepackage{amsmath}

\begin{document}


\hsize\textwidth\columnwidth\hsize\csname@twocolumnfalse\endcsname

\title{Spin-to-Charge Conversion of Mesoscopic Spin Currents}
\author{Peter Stano$^{1,2}$ and Philippe Jacquod$^{1,3}$
\\ \textit{$^1$Physics Department, University of Arizona, 
1118 East Fourth Street, Tucson, Arizona 85721, USA\\
$^2$Institute of Physics, Slovak Academy of Sciences, Bratislava 845 11, Slovakia\\
$^3$College of Optical Sciences, University of Arizona, 1630 East University Boulevard, Tucson, Arizona 85721, USA}}


\vskip1.5truecm
\begin{abstract}
Recent theoretical investigations have shown that 
spin currents can be generated by passing electric currents 
through spin-orbit coupled mesoscopic systems. Measuring these 
spin currents has however not been achieved to date. We show how
mesoscopic spin currents in lateral heterostructures can be measured 
with a single-channel voltage probe. In the presence of a spin
current, the charge current $I_{\rm qpc}$ 
through the quantum point contact connecting the probe is odd in an
externally applied Zeeman field $B$, while it is even
in the absence of spin current. Furthermore, the zero-field derivative $\partial_B I_{\rm
qpc}$ is proportional to the magnitude of the spin
current, with a proportionality coefficient that can be 
determined in an independent measurement. We confirm these findings numerically. 
\end{abstract}
\maketitle

{\it Introduction.} One of the main challenges of semiconductor spintronics is to convert hardly accessible
spin currents and accumulations into easily measured
electric currents or voltages~\cite{fabian2007:APS}.
While in metals, this challenge
is rather successfully met by means of ferromagnetic detectors~\cite{jedema2002:N}, 
uncovering spin currents and accumulations in lateral 
semiconductor heterostructures is significantly harder, because 
ferromagnets do not connect well to two-dimensional electron gases.
Instead, one uses in-plane magnetic fields that couple dominantly to 
the spin of the electrons. Thanks to the resulting Zeeman field a quantum point contact (QPC) has been polarized~\cite{thomas1996:PRL} and spin orientations in 
few-electron quantum 
dots~\cite{elzerman2004:N,shorubalko2007:N,amasha2008:PRB},
spin currents flowing out of Coulomb blockaded
quantum dots~\cite{potok2003:PRL} and spin currents 
injected from a polarized 
point contact~\cite{potok2002:PRL,koop2008:PRL,frolov2009:PRL} 
have been converted into electrostatic voltages.
In all these instances, large magnetic 
fields $B \gg 1$ T are required 
both for generating and measuring spins. These protocols are therefore
not viable for measuring independently generated spin currents --
such as, for instance, the theoretically predicted magnetoelectric mesoscopic 
spin currents~\cite{kiselev2001:APL,kiselev2003:JAP,bardarson2007:PRL,nazarov2007:NJP,krich2008:PRB,adagideli2010:PRL,ren2006:PRL} -- because 
the latter are unavoidably 
modified by such large Zeeman fields~\cite{zumbuhl2004:PRB}. 

\begin{figure}
\centerline{\psfig{file=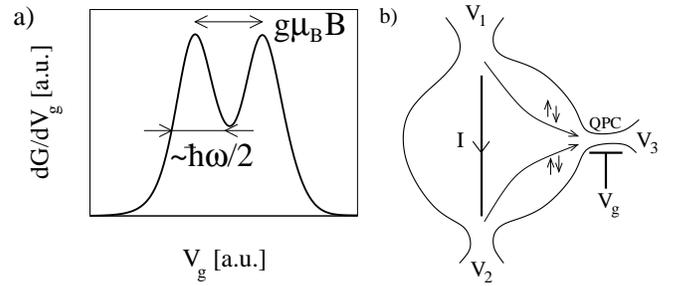,width=1\linewidth}}
\caption{(a) QPC transconductance
$d G/d V_g$ at a large magnetic field $B\gtrsim 5$ T showing how the ratio of the $g$-factor and 
the QPC energy resolution $\hbar \omega$ in Eq.~(\ref{eq:main result}) can be determined.
(b) Proposed setup for measuring mesoscopic spin currents.
A voltage probe is connected to a two-terminal lateral 
quantum dot via a gate-defined (with gate potential $V_g$) single-channel
QPC. The spin current through the QPC is converted into an electric
signal by applying a sub-Tesla in-plane magnetic field.
}
\label{fig:scheme}
\end{figure}

In this manuscript, we propose a novel scheme to 
measure mesoscopic spin currents. 
The basic principle of our proposal
is that a pure spin current flowing through a QPC results
in an odd dependence of the charge current $I_{\rm qpc}(B)$ on an externally applied Zeeman field $B$. 
Setting the voltage behind the QPC such that $I_{\rm qpc}(B=0)=0$,
the zero-field derivative $\partial_B I_{\rm qpc}\vert_{B=0}$
is proportional to the spin current at $B=0$, with a proportionality
coefficient given by the ratio of the $g$-factor and the energy resolution
of the QPC. This prefactor can be extracted independently,
either at a large magnetic field, as sketched in Fig.~\ref{fig:scheme}(a), or determining the QPC transconductance width at $B=0$ if the $g$-factor is known.
Thus, in our scheme the spin current can be {\it quantitatively} determined by measuring an electric signal. 
The scheme works in multiterminal setups, such as the one
sketched in Fig.~\ref{fig:scheme}(b), which are free
of Onsager reciprocity relations~\cite{onsager1931:PR,buttiker1986:PRL},
since the latter impose $I_{\rm qpc} (B)=I_{\rm qpc} (-B)$ in two-terminal
geometries. For a few-micron quantum dot in $n$-doped GaAs, we estimate
a signal of 10 pA in a field $B \simeq 0.5$ T, for which
currents are only weakly different from their zero-field 
value~\cite{zumbuhl2004:PRB} and the QPC is far from polarization. 
Because $I_{\rm qpc}(B=0)=0$, this signal is well above the current 
detection threshold. The scheme works at smaller fields in
materials with larger spin-orbit coupling such as $p$-type 
GaAs~\cite{grbic2007:PRL}, which are expected to carry larger spin currents.

{\it Geometry and main result.}  While our measurement scheme is rather general
and in particular works independently
of the source of spin current, we focus on
a three-terminal ballistic quantum dot 
as shown in Fig.~\ref{fig:scheme}(b). An electric current is driven by
a voltage bias $V=V_2-V_1$ applied between 
terminals one and two. A third terminal is connected to the dot through a QPC. 
The terminals
carry $N_{1,2} \gg 1$ and $N_3=1$ spin-degenerate transport channels.
We assume that spin-orbit interaction is strong enough that the spin-orbit
time is shorter than the electronic dwell time inside the dot. Spin rotational symmetry is then totally
broken and the charge current is generically accompanied by spin currents
flowing through each terminal, with a typical magnitude 
$I_{1,2,3}^{(\alpha)} = {\cal O}
(e^2V/h)$~\cite{ren2006:PRL,bardarson2007:PRL,nazarov2007:NJP,krich2008:PRB}. {Note we use $\alpha \in \{x,y,z\}$ for spin and $\alpha=0$ for charge quantities, respectively.}
Our goal is to measure the spin current through terminal three. 
To that end, the terminal is initially a
voltage probe, with 
$V_3$ and the gate potential $V_g$ defining the QPC set such that no current flows, $I_3^{(0)}=0$, and the QPC
transmission $\Gamma=G/(2e^2/h)$ is about one half, with the QPC conductance $G$. 
As we will show below, the spin current through terminal three
can be converted into an electric signal when an in-plane magnetic
field is applied. Our main result is the relation
\begin{equation}
I_3^{(\alpha)}(B=0) \simeq \frac{\hbar \omega}{\pi \mu} \partial_B I_3^{(0)}|_{B=0}
 \, ,
\label{eq:main result}
\end{equation}
between the spin current $I_3^{(\alpha)}$
in the direction $\alpha$ along the magnetic field ${\bf B}$ and the zero-field derivative of the 
charge current $I_3^{(\alpha=0)} \equiv I_{\rm qpc}$. Here, $\mu = g \mu_{\rm B}/2$ is the effective magneton in the dot's material and 
$\hbar \omega$ gives the QPC energy resolution. Both quantities can 
be extracted independently of the measurement of the spin current,
by looking at the QPC transconductance $d G/d V_g$ [see Fig.~\ref{fig:scheme}(a)]. 
It is therefore possible to
{\it quantitatively} measure spin currents. We are unaware of other
proposals for such quantitative measurement. 

{\it Scattering approach to transport.} We briefly sketch our theory.
In linear response, 
charge and spin currents in the terminals are related to voltages
by the relation~\cite{bardarson2007:PRL,buttiker1986:PRL}
\begin{equation}
I_i^{(\alpha)} = \frac{e^2}{h} \sum_{j} \left( 2 N_i \delta_{ij} \delta_{\alpha 0} - \mathcal{T}_{ij}^{(\alpha)} \right) V_j \, ,
\label{eq:current definition}
\end{equation}
assuming no spin accumulation in the leads. 
The generalized transmission coefficients are given by
\begin{equation}
\mathcal{T}_{ij}^{(\alpha)} = \sum_{m\in i, n\in j} {\rm Tr} \left( t_{mn}^\dagger \sigma^{(\alpha)}  t_{mn}  \right),
\label{eq:transmissions}
\end{equation}
with Pauli spin matrices $\sigma^{(\alpha)}$ ($\sigma^{(0)}$ is the identity
matrix).
The $2 \times 2$ matrices $t_{mn}$ are transmission 
elements of the scattering matrix. They alternatively define the 
transmission probabilities $T_{ij}^{\sigma \sigma^\prime}=\sum_{mn} 
|(t_{mn})_{\sigma\sigma^\prime}|^2 $ for an 
electron with spin $\sigma^\prime$ impinging in channel $n$ 
of terminal $j$ to exit in channel $m$ of terminal $i$ with spin $\sigma$. 

From now on we focus our discussion on the QPC charge current
$I_3^{(0)}$ and spin current $I_3^{(\alpha)}$.
To incorporate the QPC and its $B$-dependence into our theory, we make
the following two assumptions. First, we model the QPC 
as a spin-diagonal $2 \times 2$ matrix, whose elements 
depend only on the Zeeman energy $\pm \mu B$, thus on the
spin of exiting electrons. Accordingly, we write
\begin{equation}
T_{3i}^{\sigma \sigma^\prime} (B)\approx 
\tau_{3i}^{\sigma \sigma^\prime} (B) \Gamma(E_F-\sigma \mu B) \, , \qquad i=1,2,
\label{eq:central}
\end{equation}
with the transmission $\tau_{3i}^{\sigma \sigma^\prime}$ defined when the QPC 
fully transmits both spin species. 
Equation \eqref{eq:central} is valid if, upon reflection from the probe, the electron has a negligible probability to come back to the probe again. This condition is satisfied for $N_1+N_2\gg N_3$. The validity of Eq.~\eqref{eq:central} is confirmed by our numerical results, where it does not enter.

The QPC transmission $\Gamma$ is a function of the particle's kinetic energy,
with $\sigma=\pm$ for spins aligned (antialigned) with $B$. We take the standard expression~\cite{buttiker1990:PRB}
\begin{equation}
\Gamma(E_F)=\left\{1 + \exp[-2 \pi (E_F-V_g)/\hbar \omega ] \right\}^{-1} \, ,
\label{eq:QPC}
\end{equation}
with the gate voltage $V_g$ defining the QPC and $\hbar \omega$ its energy resolution.
The exact form of $\Gamma(E)$ is unimportant.
Second, we assume that the QPC has a high
sensitivity to the Zeeman field, so that in Eq.~(\ref{eq:central}), 
$\Gamma(E_F-\sigma \mu B)$ varies faster
than $\tau_{3i}^{\sigma \sigma^\prime} (B)$ with $B$.

The condition that $I_3^{(0)}(B=0) = 0$ translates into 
$V_3=(\mathcal{T}_{31}^{(0)} V_1 + \mathcal{T}_{32}^{(0)} V_2)/
(\mathcal{T}_{31}^{(0)} + \mathcal{T}_{32}^{(0)})$. 
The spin current reads
\begin{equation}
I_3^{(\alpha)} = \frac{e^2}{h} \left[\mathcal{T}_{31}^{(\alpha)} (V_3-V_1) 
+ \mathcal{T}_{32}^{(\alpha)} (V_3-V_2) \right] \, ,
\label{eq:spincurrent}
\end{equation}
and the zero-field derivative of the electric current is
\begin{equation}
\partial_B I_3^{(0)}\vert_{B=0} = \frac{e^2}{h} \left(\partial_B \mathcal{T}_{31}^{(\alpha)}\vert_{B=0} V_1 
+ \partial_B \mathcal{T}_{32}^{(\alpha)}\vert_{B=0} V_2 \right) \, .
\label{eq:currentderivative}
\end{equation}
Combining Eqs.~(\ref{eq:central}-\ref{eq:currentderivative}) directly
gives Eq.~(\ref{eq:main result}).

When $V_g=E_F$ we write $\Gamma(E_F-\sigma \mu B)
=1/2 - \sigma \gamma(B)$ and straightforwardly obtain
\begin{eqnarray}
I_3^{(0)} & = & \frac{e^2}{h} \left\{[\tau_{31}^{(0)}(B) \Gamma(0)+
\tau_{31}^{(\alpha)}(B) \gamma(B)] (V_3-V_1) \right. \nonumber \\
&& \left.
+ [\tau_{32}^{(0)}(B) \Gamma(0)+
\tau_{32}^{(\alpha)}(B) \gamma(B)]
(V_3-V_2) \right\} \! .
\end{eqnarray}
We defined $\tau_{ij}^{(0)}=\sum_{\sigma\sigma^\prime}\tau_{ij}^{\sigma\sigma^\prime}$ and  $\tau_{ij}^{(\alpha)}=\sum_{\sigma\sigma^\prime}\sigma \tau_{ij}^{\sigma\sigma^\prime}$. 
In the absence of spin-orbit interaction, and assuming that 
$B$ has no orbital effect, $B \leftrightarrow -B$ amounts to 
interchanging ``+'' and ``-'' spin directions along $\alpha$, in
which case $\tau_{3i}^{(0)}(B)=\tau_{3i}^{(0)}(-B)$, but
$\tau_{3i}^{(\alpha)}(B)=-\tau_{3i}^{(\alpha)}(-B)$ for $\alpha \ne 0$.
Noting that $\gamma(B)$ is an odd function of $B$ we conclude
that in the absence of spin-orbit interaction, hence of spin current at
$B=0$, $I_3^{(0)}$ is even in $B$. A similar conclusion is reached
for a two-terminal geometry for which $\tau_{ij}^{(0)}(B)=\tau_{ji}^{(0)}(B)
=\tau_{ij}^{(0)}(-B)$~\cite{onsager1931:PR,buttiker1986:PRL}. This corroborates
the conclusion of Ref.~\cite{adagideli2006:PRL}, that conductance measurements
at magnetic fields of opposite directions cannot access spin currents in two-terminal geometries.

{\it Numerical model and results.} Having discussed our theory, we now illustrate it 
numerically. We consider a two-dimensional quantum dot 
in the single band effective mass approximation.
The Hamiltonian for conduction electrons reads
\begin{equation} 
H = \frac{ {\bf p}^2 }{2 m} + v({\bf r}) + \mu {\bf B} \cdot \boldsymbol{\sigma} + \frac{\hbar}{2 m l_{\rm br}} (\sigma_x p_y-\sigma_y p_x) \, ,
\label{eq:hamiltonian}
\end{equation}
with the electron effective mass $m$, the momentum 
operator ${\bf p}=-{\rm i}\hbar \boldsymbol{\nabla}$, 
the in-plane magnetic field ${\bf B}$, with $|{\bf B}|=B$, 
and the vector $\boldsymbol{\sigma}$ of Pauli matrices. We specified to 
Bychkov-Rashba spin-orbit interaction, parametrized by the spin-orbit 
length $l_{\rm br}$, but stress that our theory is equally valid for other 
forms of spin-orbit interaction. 
The potential $v({\bf r})$ models both the dot's hard wall confinement and
a smooth disorder inside the dot. The latter is tailored to minimize
direct transmission from lead to lead and make our numerics as generic as
possible. 

We take leads as semi-infinite waveguides, without
spin-orbit interaction. Spin currents in the leads are then well 
defined~\cite{rashba2003:PRB}. 
The QPC is modeled as a narrowing of the dot towards the third lead 
through an inverted parabolic potential
\begin{equation}
v_{\rm QPC}({\bf r}^\prime)= V_g - m \omega^2 x^{\prime 2}/2 \, .
\label{eq:QPC potential}
\end{equation}
Here the primed coordinate is measured from the QPC center, $V_g$ is the gate 
potential used to tune the QPC transmission, and $\hbar \omega$ sets the QPC 
energy resolution. These parameters are model dependent, but their ratio has 
a clear experimental meaning in terms of the $B$-field response
of the QPC transconductance $dG/dV_g$. This
is illustrated in Fig.~\ref{fig:scheme}(a). Equation (\ref{eq:QPC potential}) is
consistent with the transmission given in Eq.~(\ref{eq:QPC}). As argued above, our 
measurement scheme works best when the QPC is most sensitive
to energy variations and accordingly we set its potential at
the Fermi energy, $V_g=E_F$. 

\begin{figure}
\centerline{\psfig{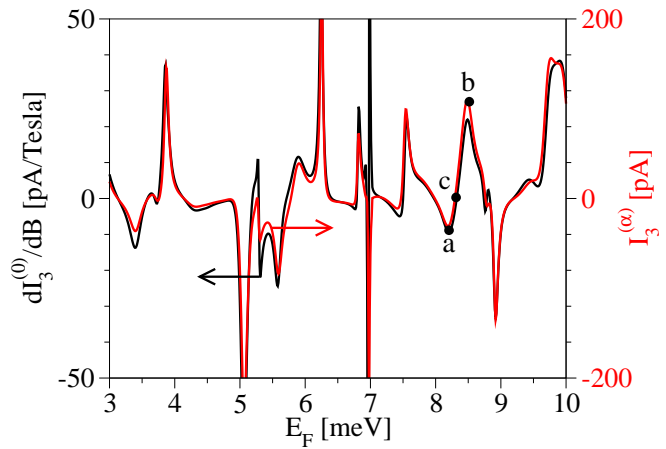}}
\caption{(Color online) Comparison of the spin current $I_3^{(\alpha)}(B=0)$
(red or gray line; right y axis) and the zero-field derivative 
$\partial_B I_3^{(0)}\vert_{B=0}$ of the charge current
(black line; left y axis) measured at terminal 3, as a function
of the Fermi energy $E_F$. The
tags a-c refer to panels in Fig.~\ref{fig:even odd}.}
\label{fig:proof}
\end{figure}

We use material parameters corresponding to GaAs heterostructures, i.e.
$m=0.067 m_e$, with $m_e$ the free electron mass, $g=-0.44$, and 
we vary $E_F\in[3,10]$ meV. Leads 1 and 2 are 50 nm wide, corresponding
to up to three open transport channels in the considered energy range.
The terminal voltages are $V_1=-V_2=50$ $\mu$V and $V_3$ is set 
such that $I_3^{(0)}(B=0)=0$. For the QPC we set 
$\hbar \omega \approx 0.18$ meV, corresponding to a
spin resolution at $B\approx 6$ T,
and we take it to be 1.2 $\mu$m long to obtain numerically 
sharp conductance steps. The temperature effects on the QPC transmission 
can be neglected if $k_B T \ll \hbar \omega$, which is 
fulfilled at sub-Kelvin temperatures. 
Our numerics limits the dot
linear size to about 200 nm, and accordingly we scale down the spin-orbit length
to $l_{\rm br}=100$ nm, about an order of magnitude stronger than is typical 
for GaAs heterostructures, where 10 times larger dots have broken
spin rotational symmetry~\cite{folk2001:PRL}.
As an unwanted numerical artifact the QPC itself affects the spin, being 
much longer than the spin-orbit length. 
These effects are minimal
in the presence of only one type of spin-orbit interaction and we choose the Bychkov-Rashba one. We 
checked, but do not show that our numerical results are qualitatively unchanged for other
types of spin-orbit interaction.

We first illustrate the validity of 
Eq.~(\ref{eq:main result}) in Fig.~\ref{fig:proof}. We see
that the zero-field derivative of the charge current in lead 3 faithfully
follows the spin current, despite large fluctuations of the latter 
as the Fermi energy is varied. The two quantities are almost perfectly
correlated, except close to 7 meV, where the number of channels in 
leads 1 and 2 artificially jumps from 2 to 3 due to the way we
model the leads. This is a numerical artifact. We numerically
calculated the current derivative as $\partial_B I_3^{(0)}\vert_{B=0} = 
[I_3^{(0)}(B/2)-I_3^{(0)}(-B/2)]/B$ with $B=10^{-3}$ T. 
Experimentally, however, the magnetic field must be large enough that
the current change is measurable, but still small enough that
(i) it does not generate mesoscopic fluctuations of the transmission 
coefficient $\tau_{3i}^{\sigma \sigma^\prime} (B)$ 
[see Eq.~(\ref{eq:central})], (ii) it does not polarize the QPC, since
this would make $I_3^{(\alpha)}$ saturate, and (iii) it does not
freeze spin-orbit interaction inside the dot. The upper bound on $B$ comes 
from (i) since, according to Ref.~\cite{zumbuhl2004:PRB}, 
$\tau_{3i}^{\sigma \sigma^\prime} (B)$ decorrelates at a field of
about 1 T for a ballistic micron sized GaAs dot, while 
bounds on (ii) and (iii) are at $B \gtrsim 5$ T or more.
Limiting ourselves to fields of 0.5 T, we estimate 
from Fig.~\ref{fig:proof} 
that a current sensitivity of about 10 pA is sufficient for spin-to-charge
conversion of typical spin currents in ballistic lateral dots in GaAs.

\begin{figure}
\centerline{\psfig{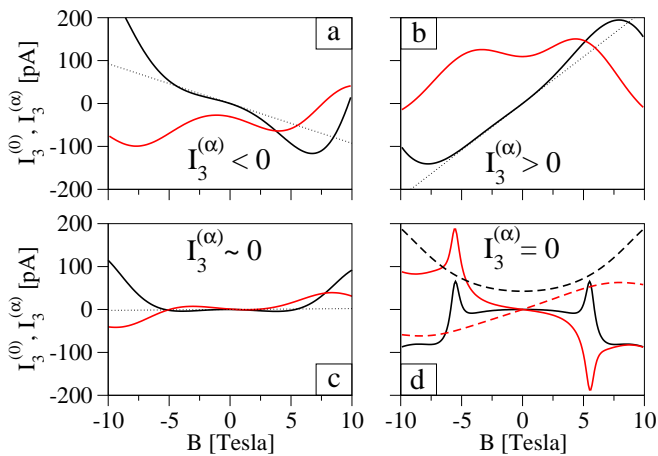}}
\caption{(Color online) Spin (red or gray lines) and charge (black lines) currents in terminal 3 as a function of the in-plane magnetic field $B$. (a)-(c) Spin-orbit coupled dots corresponding to the labeled data points 
in Fig.~\ref{fig:proof}. Dotted lines show the slope of the charge current at $B=0$. (d) Dot without spin-orbit interaction (solid lines) and dot with voltages $V_1=V_2=V/2=-V_3$ (dashed lines).}
\label{fig:even odd}
\end{figure}

We finally focus on the parameter sets for the data points
labeled ``a'', ``b'' and ``c'' in Fig.~\ref{fig:proof},
corresponding to  negative, positive, and zero spin current at $B=0$
respectively. The first three panels of Fig.~\ref{fig:even odd} show the
magnetic-field dependence of $I_3^{(0)}$ and $I_3^{(\alpha)}$ in these
three instances. The data clearly illustrate that the sign and
magnitude of the spin current at $B=0$ is reflected in the 
slope of the electric current. We furthermore see that the electric 
current is linear up to magnetic field of 1-2 T, up to 
where, therefore, the
zero-field derivative of the electric current can still be extracted. 
Figure \ref{fig:even odd}(d) additionally shows that the current is exactly even in $B$
in the absence of spin current. This would happen in the absence of spin-orbit interaction, or if the leads 1 and 2 are set to the same voltage, biased with respect to the single-channel lead 3. The latter case provides a simple check of our method, as in this setup the spin current is forbidden~\cite{zhai2005:PRL,kiselev2005:PRB}. 

While we focused on a QPC set to a maximal sensitivity, $\Gamma\vert_{B=0}=1/2$,
our theory remains valid away from there [or for a QPC with a transmission different from the one in Eq.~\eqref{eq:QPC}] provided one substitutes 
$\pi \mu/\hbar \omega \to \partial_B {\rm ln}\Gamma|_{B=0}$ in Eq.~(\ref{eq:main result}).
Also, we considered $V_3$ fixed while changing B. An alternative is to set it such that $I_3^{(0)}(B)=0$. Then Eq.~(\ref{eq:main result}) is replaced by 
\begin{equation}
I_3^{(\alpha)}(B=0) \simeq \frac{\hbar \omega}{\pi \mu} \frac{e^2}{h} [2-\mathcal{T}_{33}^{(0)}] \partial_B V_3(B)|_{B=0}\, .
\end{equation} 
The spin current can thus also be extracted from a voltage measurement, however, this additionally requires a measurement of $\mathcal{T}_{33}^{(0)}$.

{\it Conclusions.} Our theoretical and numerical investigations
show how mesoscopic spin currents can be converted into electric signals
by measuring the magnetic-field response 
of the electric current through a QPC. Qualitatively, the presence or
absence of a spin current is directly reflected in the symmetry of the 
electric current through the QPC. 
We moreover demonstrated that, beyond emphasizing the presence of
a spin current, our measurement scheme renders  
the magnitude of the current quantitatively accessible, since the proportionality 
coefficient in Eq.~(\ref{eq:main result})
can be experimentally extracted from the transconductance of the
QPC at a large Zeeman field.
We estimate that typical spin currents flowing in GaAs
quantum dots with broken spin rotational symmetry have a measurable
electric signature at magnetic fields that are low enough that the
targeted spin current is not altered by the measurement process. 
Finally, we stress that our scheme works independently of the source of spin current.

{\it Acknowledgements.} We would like to thank Brian LeRoy for valuable discussions. This work has been supported by the NSF under Grant No.~DMR-0706319.

\bibliography{../references/quantum_dot}

\end{document}